\documentclass[showpacs,floatfix,aps,prl,preprint,superscriptaddress]{revtex4-1}
\usepackage{float}
\usepackage{graphicx}
\usepackage{amsmath}
\usepackage{epstopdf}
\usepackage[caption=false]{subfig}
\usepackage{xfrac}
\usepackage{amssymb}
\usepackage{braket}
\usepackage{graphicx}
\usepackage{verbatim}
\usepackage{etoolbox}
\usepackage{amsfonts} 
\usepackage{amsmath} 
\usepackage[utf8]{inputenc} 
\usepackage{mathtools}
\usepackage{soul,xcolor,colortbl}
\usepackage{lipsum}
\usepackage{hyperref} \hypersetup{colorlinks=true,citecolor=blue,linkcolor=blue,urlcolor=blue} 

\usepackage{xcolor, amssymb} 

\usepackage{ulem}
\usepackage{bm} 
\usepackage{array}
\usepackage{color}

\begin{document}


\title{The Dislocation Content of Triple Junctions}

\author{Ian S. Winter}
\email{iswinte@sandia.gov}
\affiliation{Sandia National Laboratories, 7011 East Ave., Livermore, CA 94550 USA}
\author{R. Daniel Moore}
\affiliation{Sandia National Laboratories, 7011 East Ave., Livermore, CA 94550 USA}
\author{R. E. Rudd}
\affiliation{Lawrence Livermore National Laboratory, 7000 East Ave., Livermore, CA 94550 USA}
\author{T. Oppelstrup}
\affiliation{Lawrence Livermore National Laboratory, 7000 East Ave., Livermore, CA 94550 USA}
\author{T. Frolov}
\email{frolov2@llnl.gov}
\affiliation{Lawrence Livermore National Laboratory, 7000 East Ave., Livermore, CA 94550 USA}

\date{\today}

\begin{abstract}

Triple junctions, line defects formed by the intersection of different grain boundaries, exist within all polycrystalline materials. While it has long been recognized that triple junctions could play an important role in microstructural evolution, there remains much uncertainty regarding their properties. Triple junctions are line defects capable of carrying dislocation content. However, no general method for calculating this content has been established. In this work, we derive the necessary equations to calculate the intrinsic dislocation content of a triple junction whose trichromatic pattern forms a coincidence site lattice. We further show that this approach can be easily applied to facet junctions, and in principle, any type of grain boundary junction for which a coincidence site lattice can be defined. We apply this formalism to atomistic simulations of tungsten to compute the Burgers vectors of a facet junction and a triple junction formed during twin grain nucleation and growth from a free surface. By tracking the evolution of the triple junction's Burgers vector and its core structure, we reveal the sequence of individual line defect reactions responsible for triple-junction-mediated twin growth.

\end{abstract}

\maketitle

\section{Introduction}

Triple junctions (TJs), line defects formed by the intersection of three grain boundaries, necessarily exist in all polycrystalline materials. Classically, triple junctions were described purely in terms of the grain boundaries that form them, such as through the Herring condition \cite{Herring}. However, it has long been recognized that triple junctions have structure and properties which are distinct from both the bulk phases and the grain boundaries that form them \cite{FORTIER1991177,ZHAO20105646,KIM20093662,King01052007,upmanyu1999triple,UPMANYU20021405,SRINIVASAN19992821,EICH2016364,TUCHINDA2024120274,TUCHINDA2025121429}. For example, experimental measurements of impurity diffusion suggested that diffusivity can be three orders of magnitude faster along TJs compared to GBs \cite{CHEN2007253}. The assumption of fast TJ diffusion  was used to explain the anomalous diffusion and creep behavior of nanocrystalline materials \cite{CHEN2007253,FEDOROV200251}. Atomistic simulation studies of self-diffusion, on the other hand, showed that TJ diffusivity is comparable to the diffusivity of high-angle grain boundaries \cite{PhysRevB.79.174110}. TJs were also found to be preferred sites of solute segregation for a range of systems \cite{Barnett2024,PENG2022117522,REDACHELLALI2013164,Yin2003}. For example, both experiments and simulations indicated strong Au segregation to TJs in a nanocrystalline Pt-Au alloy, which pinned grain boundaries and suppressed grain growth \cite{Barnett2024}. Computational studies of solute segregation in binary alloys calculated TJ segregation spectra and concluded that solutes show either a preference for the GBs or TJs depending on the system \cite{TUCHINDA2025121429}.

To reflect the distinct contribution of TJs to properties of polycrystalline materials, early phenomenological models treated TJs as triple lines in fluids; assigning excess free energy per unit length and a distinct mobility \cite{CZUBAYKO19985863,GOTTSTEIN2002703,GOTTSTEIN20051535,JOHNSON2014134}. In this treatment, the excess energy of a TJ was attributed solely to its core and therefore scaled linearly with its length. These capillary models were used to predict how triple junctions modify the capillary contact equilibrium conditions, describe the role of triple junctions in interactions between grain boundaries and precipitates, Zener pinning, and back out triple junction energy values from experiments \cite{King01052007,KIM20093662,ZHAO20105646} and simulations \cite{SRINIVASAN19992821,EICH2016364,TUCHINDA2024120274,TUCHINDA2025121429}. These models successfully explained different experimentally observed grain growth regimes: linear growth at low temperatures where TJ drag was expected to dominate, and parabolic growth at high temperatures suggested to be controlled by curvature-driven boundary migration \cite{CZUBAYKO19985863,Gottstein1999,GOTTSTEIN2000397,GOTTSTEIN2002703,Upmanyu1999,UPMANYU20021405,JOHNSON2014134}. However, these models did not account for the possible dislocation character of TJs and their system-size-dependent energy due to their long-range elastic field.

A more recent disconnection-based theory of TJ motion recognizes the discrete nature of TJ migration governed by the fluxes of disconnections flowing into and out of the junction from its three constituent boundaries \cite{PhysRevLett.119.246101,Thomas2019,HAN2022117178}. In this picture, the Burgers vector content of the TJ changes during migration as incoming disconnection fluxes must satisfy both a geometric connectivity condition and a Burgers vector balance. Consequently, TJ mobility is not an intrinsic scalar property, but an emergent quantity governed by disconnection availability and temperature, naturally accounting for both drag-dominated and fast-migration regimes. While this theory emphasizes the importance of dislocation defects in triple junction migration, it does not consider  the intrinsic Burgers content of triple junctions.

 It has been shown using circuit mapping \cite{POND1994287} that a general TJ can have dislocation content of its own, independent of dislocations or disconnections absorbed at the TJ \cite{dimitrakopulos1997defect}. While the work in Ref. \cite{dimitrakopulos1997defect} shows that TJs can have dislocation content due to the topology of the defect, it does not give a prescription for how to calculate that value.  In this work, we utilize a recently developed formalism that describes the microscopic degrees of freedom of an individual grain boundary in terms of a single vector, $\bm{t}^{WS}$ located inside the Wigner–Seitz cell of the displacement-shift-complete lattice (DSCL) \cite{WINTER2025120968}. By knowing $\bm{t}^{WS}$ of the three grain boundaries that constitute a TJ, we show that we can explicitly define all possible values of the Burgers vector that can exist at the TJ.

  The remainder of this paper is organized as follows. First, we derive the general Burgers circuit formalism for quantifying the dislocation content of a grain boundary junction. We show that the Burgers vector of any junction whose constituent grains share a coincidence site lattice can be expressed in closed form in terms of the $\bm{t}^{WS}$ vectors \cite{WINTER2025120968} of the three constituent boundaries and a vector of the DSCL. We further demonstrate that the same framework applies naturally to facet junctions, and, by extension, to any GB junction for which a DSCL can be defined. We apply this formalism to atomistic simulations of tungsten, computing the dislocation content of both a facet junction and a triple junction formed during the nucleation and growth of a grain on a free surface. 
Burgers circuit analysis performed during the growth allows us to track the evolution of the Burgers vector and core structure of the triple junction during its motion and reveals the sequence of individual line defect reactions responsible for triple junction mediated twin growth. The thermodynamic implications of the long-range elastic field associated with TJ dislocation content are discussed.

\section{Theory} \label{sec:theory}

\begin{widetext}
\begin{figure}[ht!]
\centering
\includegraphics[scale=1]{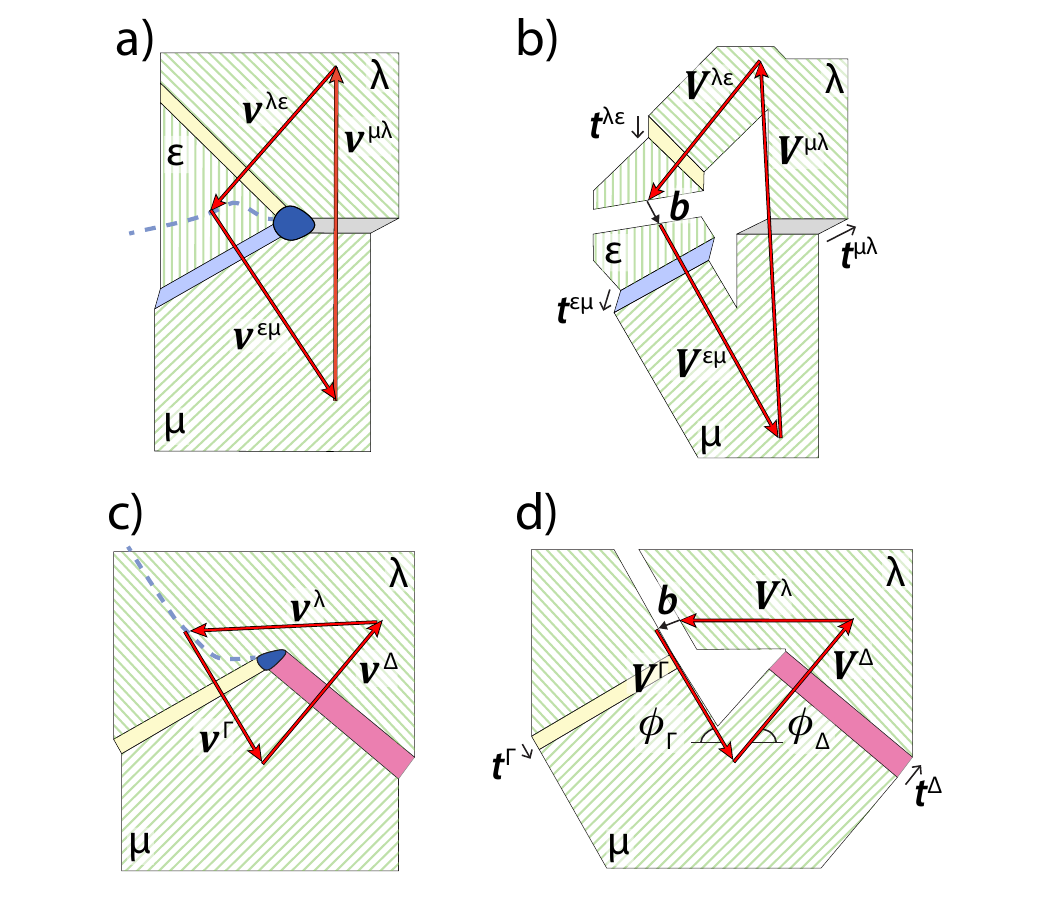}
\caption{
The Burgers circuit used to calculate the Burgers content of a GB triple junction, panels a and b; facet junction, panels c and d. Panels a and c show the Burgers circuit in the current system for each junction, with the dark blue region signifying the core of the dislocation associated with the junction. The light blue, yellow, gray and pink regions represent the respective grain boundaries. Panels b and d show the Burgers circuit in the reference system. The closure failure, and its associated Burgers vector, can be clearly seen in panels b and d. The cut associated with the creation of the reference systems is represented as a dashed line and is displayed in panels a and c.} 
\label{fig:burgers_circuit_general_junction}
\end{figure}
\end{widetext}

Let us consider a GB triple junction such as the general one shown in Fig. \ref{fig:burgers_circuit_general_junction}a. We draw a closed circuit around the triple junction. The circuit is composed of three crossing vectors across the three grain boundaries that make up the triple junction. Let each crossing vector connect a lattice site in one grain to a lattice site in another grain. We then define a cut, such as the blue dashed line shown in Fig. \ref{fig:burgers_circuit_general_junction}a that does not intersect any of the crossing vectors. The cut allows the system to relax to a reference system that is composed of three independent grain boundaries as shown in Fig. \ref{fig:burgers_circuit_general_junction}b. 

As discussed in Ref. \cite{WINTER2025120968}, the structure of a grain boundary can be uniquely represented by a displacement vector, $\bm{t}^{WS}$ located inside the Wigner–Seitz cell of the DSCL. The components of $\bm{t}^{WS}$ are related to GB excess properties, including excess volume, number of atoms and excess shears. As such, each of the crossing vectors can be represented as a combination of $\bm{t}^{WS}$ and lattice vectors from each grain comprising the GB. Thus the crossing vector between the grains $\lambda$ and $\mu$ can be expressed as $\bm{V}^{\mu\lambda} = \bm{U}^{\mu}+\bm{U}^{\lambda}+\bm{t}^{WS,\mu\lambda}$, where $\bm{U}^{\lambda}$ and $\bm{U}^{\mu}$ are lattice vectors in the $\lambda$ and $\mu$ grains. Using the $RH/FS$ formalism for a Burgers circuit \cite{HirthLothe}, we can write the Burgers vector of a triple junction to be:
\begin{subequations}
    \begin{align}
        \bm{b} &= -\left(\bm{V}^{\mu\lambda}+\bm{V}^{\lambda\varepsilon}+\bm{V}^{\varepsilon\mu}\right),\label{eq:tj-circuit}\\
        \bm{V}^{\mu\lambda} &= \bm{U}^{\mu}+\bm{U}^{\lambda}+\bm{t}^{WS,\mu\lambda},\\
        \bm{V}^{\lambda\varepsilon} &= \bm{W}^{\lambda}+\bm{W}^{\varepsilon}+\bm{t}^{WS,\lambda\varepsilon},\\
        \bm{V}^{\varepsilon\mu} &= \bm{Y}^{\varepsilon}+\bm{Y}^{\mu}+\bm{t}^{WS,\varepsilon\mu},\\
        \bm{b} &= -\left(\bm{U}^{\mu}+\bm{Y}^{\mu} + \bm{U}^{\lambda} + \bm{W}^{\lambda} + \bm{W}^{\varepsilon} + \bm{Y}^{\varepsilon}\right)-\left(\bm{t}^{WS,\mu\lambda}+\bm{t}^{WS,\lambda\varepsilon}+\bm{t}^{WS,\varepsilon\mu}\right)\label{eq:b_dsc_def},
    \end{align}
\end{subequations}
where $\bm{U}$, $\bm{W}$ and $\bm{Y}$ are lattice vectors in the respective grains. It is important to note that all $\bm{t}^{WS}$ vectors in Eq. \eqref{eq:b_dsc_def} are defined within the same reference frame. As shown in Eq. \eqref{eq:b_dsc_def}, the Burgers vector is composed of a sum of lattice vectors of the three grains and a sum of the displacement vectors, $\bm{t}^{WS}$, of the three GBs. Viewing the lattice vector term in Eq. \eqref{eq:b_dsc_def}, it consists of a linear combination of lattice vectors from all grains. Provided that a coincidence site lattice (CSL) exists, the combination of different lattice vectors forms a lattice of its own, which is known as the displacement-shift complete lattice (DSCL) \cite{GRIMMER19741221,bollmann2012crystal}. As a result, Eq. \eqref{eq:b_dsc_def} shows that assuming a DSCL is formed by the three grains comprising the triple junction, we can always express the Burgers circuit of a grain boundary junction as 
\begin{equation}\label{eq:b_tj}
    \bm{b}^{TJ} = \bm{d}-\left(\bm{t}^{WS,\mu\lambda}+\bm{t}^{WS,\lambda\varepsilon}+\bm{t}^{WS,\varepsilon\mu}\right),
\end{equation}
where $\bm{d}$ can be any possible DSCL vector. As such, Eq. \eqref{eq:b_tj} defines all possible Burgers vectors that can exist at a triple junction. While there are certain high symmetry cases where it is possible for the $\bm{t}^{WS}$ vectors to cancel out, in the general case, this sum will be finite. As a result, Eq. \eqref{eq:b_tj} states that only quantized values of dislocation content can exist at the grain boundary. Conversely, if we consider the case where no DSCL common to the three grains exists, the DSCL is considered to be infinitely dense. This means that a continuum of Burgers vectors can exist at the grain boundary.

The Burgers circuit analysis for a triple junction can be applied to any type of GB junction for which a DSCL exists. As an example, the Burgers circuit analysis approach is applied to a facet junction in Fig. \ref{fig:burgers_circuit_general_junction}c and \ref{fig:burgers_circuit_general_junction}d. From the analysis of the facet junction it can be shown that the Burgers circuit can be written as 
\begin{equation}\label{eq:FJgeneral}
    \bm{b}^{FJ} = -(\bm{V}^{\Gamma}+\bm{V}^{\Delta}+\bm{V}^\lambda)
\end{equation}
Using the same approach as that described for a triple junction, all possible Burgers vectors at the junction are:

\begin{equation}\label{eq:FJtWS}
    \bm{b}^{FJ} = \bm{d}-\left(\bm{t}^{WS,\Gamma}+\bm{t}^{WS,\Delta}\right).
\end{equation}
To avoid confusion we define the different GB facets by capital Greek letters.

\section{Results}\label{sec:Results}

\begin{figure}[ht!]
\centering
\includegraphics[scale=1.0]{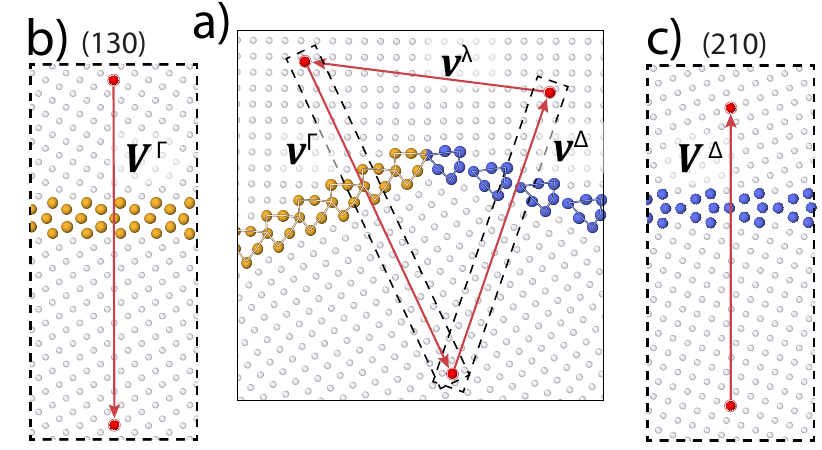}
\caption{Burgers vector calculation of a GB facet junction consisting of an intersecting $\Sigma 5 (130)[001]$  and $\Sigma 5 (210)[001]$ GBs. Panel a depicts the facet junction and the Burgers circuit around it. Panel b shows the mapping of the vector $\bm{v}^{\Gamma}$ to the reference system containing the $\Sigma 5(130)[001]$ GB only. Panel c shows the mapping of the vector $\bm{v}^{\Delta}$ to the reference system, which contains the $\Sigma 5(210)[001]$ GB only.} 
\label{fig:facet_210_310}
\end{figure}

The first example  we consider in this work is a GB facet junction that consists of the intersection of  $\Sigma 5 (130) [001]$ and $\Sigma 5 (210) [001]$ GBs as shown in Fig.\ \ref{fig:facet_210_310}, with the relevant parameters for the two GBs constituting the junction displayed in Table~\ref{tab:facet_210_310}. This junction was formed by annealing an asymmetric $\Sigma 5$ GB, with a W interatomic potential~\cite{ZHOU20014005}, up to 3000 K and then quenching to 0 K. After a conjugate-gradient minimization procedure was applied, a structure similar to that studied in Ref.~\cite{MEDLIN2017383} was created. All simulations were performed using LAMMPS~\cite{Plimpton1995Fast}, and atomic structures were visualized using OVITO~\cite{stukowski2010visualization}.

\begin{table}
\caption{\label{tab:facet_210_310} The relevant parameters for calculating the Burgers vector associated with the facet junction illustrated in Fig.\ \ref{fig:facet_210_310}. The angles $\phi$ shown in the table refer to the angles illustrated in Fig. \ref{fig:burgers_circuit_general_junction}d. $[n]$ is the GB atomic fraction.}
\begin{ruledtabular}
\begin{tabular}{llllll}
GB & $\phi\ (^{\circ}) $  & $t_1^{WS}$ ($\mathrm{\AA}$) & $t_2^{WS}$ ($\mathrm{\AA}$)  & $t_3^{WS}$ ($\mathrm{\AA}$)  & $[n]$ \\
\hline
$\Sigma 5 (210)[001]$ ($\Delta$) & $71.56$ & 0.00 & 0.00 & -0.21 & 0\\
$\Sigma 5 (310)[001]$ ($\Gamma$) & $63.43$ & 0.00 & 0.00 & 0.51 &  0
\end{tabular}
\end{ruledtabular}
\end{table}

A Burgers circuit is drawn around the facet junction as shown in Fig. \ref{fig:facet_210_310}a. The vectors $\bm{v}^{\Gamma}$ and $\bm{v}^{\Delta}$ are then mapped to the corresponding reference coordinate systems; a $\Sigma 5 (130)[001]$ and $\Sigma 5 (210)[001]$ symmetric GB. Equation \eqref{eq:FJgeneral} is then utilized to determine the Burgers content of the junction: $\bm{b} = -0.61\mathrm{\AA}\hat{\bm{e}}_1 -0.69\mathrm{\AA}\hat{\bm{e}}_3$. 

We compare the measured Burgers vector to the Burgers content due to the microscopic degrees of freedom of the constituent facets. To do so, we take Eq. \eqref{eq:FJtWS} and set $\bm{d}=\bm{0}$: $\bm{b}=-(\bm{t}^{WS,\Gamma}+\bm{t}^{WS,\Delta})$. Inserting the values from TABLE \ref{tab:facet_210_310} into Eq.\ \eqref{eq:FJtWS} we find $\bm{b}^{MDF} = -0.16 \mathrm{\AA}\hat{\bm{e}}_1 + 0.65 \mathrm{\AA}\hat{\bm{e}}_3$. This does not correspond to $\bm{b}$ found from the Burgers circuit in Fig. \ref{fig:facet_210_310}. The discrepancy between the two approaches indicates that an additional disconnection has been absorbed by the facet junction. For the system of interest, which is the DSCL associated with a $\Sigma 5$ CSL, the DSCL vectors, which can be calculated using the approach outlined in Ref. \cite{ADMAL2022118340}, are found to be

\begin{subequations}\label{eq:facet_DSC}
\begin{align}
    \bm{a}^{DSC} &= \frac{a_0}{\sqrt{5}}\left( \cos\omega\hat{\bm{e}}_1 - \sin\omega \hat{\bm{e}}_3 \right),\\
    \bm{b}^{DSC} &= \frac{a_0}{\sqrt{5}}\left( \sin\omega \hat{\bm{e}}_1 + \cos\omega\hat{\bm{e}}_3 \right),\\
    \bm{c}^{DSC} &= a_0\left(\frac{\sin\omega-\cos\omega}{2\sqrt{5}}\hat{\bm{e}}_1 + \frac{1}{2}\hat{\bm{e}}_2 -\frac{(\sin\omega+\cos\omega)}{2\sqrt{5}}\hat{\bm{e}}_3\right),
\end{align}
\end{subequations}

\noindent where $\omega$ is the angle between the $[210]$ and $[110]$ directions and $\omega \approx 18.4^{\circ}$. The 0 K lattice parameter for this system is $a_0=3.16\ \mathrm{\AA}$. With the DSCL vectors in hand, it is found that the difference, up to the significant figures given, between Eq. \eqref{eq:FJgeneral} and \eqref{eq:FJtWS} is  $\bm{b}^{DSC}=0.45\mathrm{\AA}\hat{\bm{e}}_1 + 1.34 \mathrm{\AA}\hat{\bm{e}}_3$.  

\begin{figure}[ht!]
\centering
\includegraphics[scale=0.9]{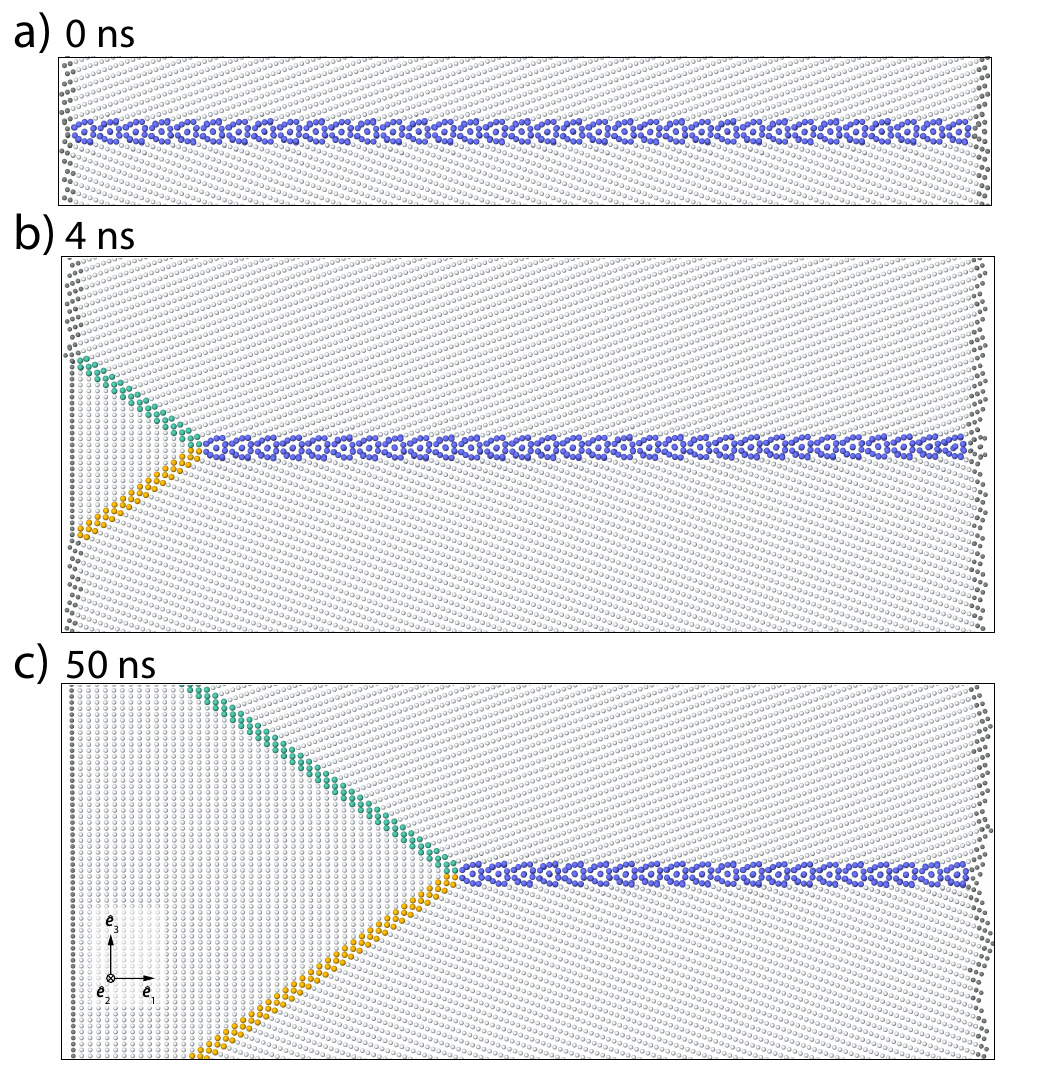}
\caption{(a) Initial configuration prior to annealing. Twin nucleation and growth simulation after a runtime of (b) 4 ns and (c) 50~ns followed by quenching. Atoms are colored to distinguish the $\Sigma$9 (blue), $\Sigma$3 (green/orange), free-surface (black), and bulk (white) atoms. Twin nucleation only occurs on the left free surface.
} 
\label{fig:twin-nucleation}
\end{figure}

For the next example, we consider the case of a symmetric $\Sigma 9 (221)[1\bar{1}0]$ tilt GB in W, using the interatomic potential from Ref. \cite{ZHOU20014005}, with open surfaces whose normals are perpendicular to the tilt axis. The initial configuration contains a non-integer GB atomic fraction: $[n]=1/2$, and was generated using the GRand Canonical Interface Predictor (GRIP)~\cite{chen2024grand}. This GB phase exhibits a complex structure, with atoms occupying interstitial positions between the (110) planes of the adjacent grains. Its formation requires a change in the number of atoms, sampling of larger area reconstruction and as such it cannot be generated using the standard $~\gamma$-surface method. The structural units shown in Fig.~\ref{fig:twin-nucleation} closely resemble those of the previously studied $\Sigma 27 (552)[1\bar{1}0]$ boundary from the same tilt family, which was also discovered using grand-canonical optimization and confirmed by DFT to be the ground state~\cite{frolov_2018}. Furthermore, the structure of the $\Sigma 9 (221)[1\bar{1}0]$ tungsten boundary predicted in this work matches the structure of the same boundary in BCC iron observed by atomic-resolution electron microscopy ~\cite{seki2023incommensurate}. Specifically, we set the maximum replications to 3 in both the $x$ and $y$ directions (GB plane), used a $z$ cutoff of 35~\AA, set the minimum and maximum temperatures to 1000~K (0.2T$_m$) and 3000~K (0.7T$_m$), respectively, used a initial gap between slabs of 0.3~\AA, and set the minimum and maximum numbers of MD steps to 10,000 and 100,000, respectively. The system dimensions were approximately $250\times 9\times 400\ \mathrm{\AA^3}$. Simulations were conducted at 2500~K for 70~ns in the canonical (NVT) ensemble (videos captured during annealing, comparing systems with periodic boundary conditions and open surfaces, are provided in the supplemental material). The bottom slab was held fixed, while the top grain contained a free-floating rigid body, following a setup similar to that described in  Ref. \cite{Frolov2013}. A Langevin thermostat~\cite{brunger1984stochastic} was employed with a damping parameter of 10~fs and a timestep of 1~fs, during which time a twin nucleated from the intersection of the left surface with the grain boundary. As a result, a triple junction between two $\lbrace112\rbrace$ twin boundaries and the $\Sigma 9$ GB was formed.  To resolve the structures, snapshots obtained during annealing (such as the one shown in Fig.~\ref{fig:twin-nucleation}) were quenched including rescaling the simulation box, to account for thermal expansion, followed by conjugate-gradient energy minimization at 0~K.


In this simulation, a twin nucleates from the left surface only (Fig. \ref{fig:twin-nucleation}). As the twin is created two twin boundaries form and the free surface changes from $\lbrace114\rbrace$ to $\lbrace110\rbrace$. If a twin were to form on the right surface, it would create a $\lbrace 001\rbrace$ surface. As such, we can write out the change in total energy per unit area of $\Sigma 9 $ GB removed, $A^{\mu\lambda}$, by the growth of the twin on the left and right surfaces as

\begin{subequations}
    \begin{align}
        \frac{\Delta E^L}{A^{\mu\lambda}} &= \frac{2\gamma^{\lambda\varepsilon}}{\cos\phi^L}+2\left(\sigma^{\lbrace110\rbrace}-\sigma^{\lbrace114\rbrace}\right)\tan\phi^L-\gamma^{\mu\lambda},\label{eq:left:grain}\\
        \frac{\Delta E^R}{A^{\mu\lambda}} &= \frac{2\gamma^{\lambda\varepsilon}}{\cos\phi^R}+2\left(\sigma^{\lbrace001\rbrace}-\sigma^{\lbrace114\rbrace}\right)\tan\phi^R-\gamma^{\mu\lambda}.\label{eq:right:grain}
    \end{align}
\end{subequations}
where $\gamma^{\mu\lambda}$  is the excess free energy of the grain boundary and $\sigma^{\lbrace hkl\rbrace}$ is the free surface free energy of the $\lbrace hkl \rbrace$ plane. $\phi^L$ and $\phi^R$ are the angles formed by the twin plane and the consumed GB plane: $\phi^L\approx 35.26$\textdegree\ and $\phi^R\approx 74.21$\textdegree. Using molecular statics simulations we find $\sigma^{\lbrace110\rbrace}=2.57\ \mathrm{J/m^2}$, $\sigma^{\lbrace001\rbrace}=2.98\ \mathrm{J/m^2}$ and $\sigma^{\lbrace114\rbrace}=3.15\ \mathrm{J/m^2}$. The GB energies are given in TABLE \ref{tab:tj_221_112}. Inputting these values, along with the GB energies from TABLE \ref{tab:tj_221_112} we find $\Delta E^L/A^{\mu\lambda}=-1.81\ \mathrm{J/m^2}$: there is a driving force for the nucleation of the twin on the left surface of the crystal. As such, this simulation describes a mechanism of annealing twin formation. Applying this same methodology to the right surface we find no such driving force for twin nucleation: $\Delta E^R/A^{\mu\lambda}= 0.66\ \mathrm{J/m^2}$.

\begin{figure}[ht!]
\centering
\includegraphics[scale=1.0]{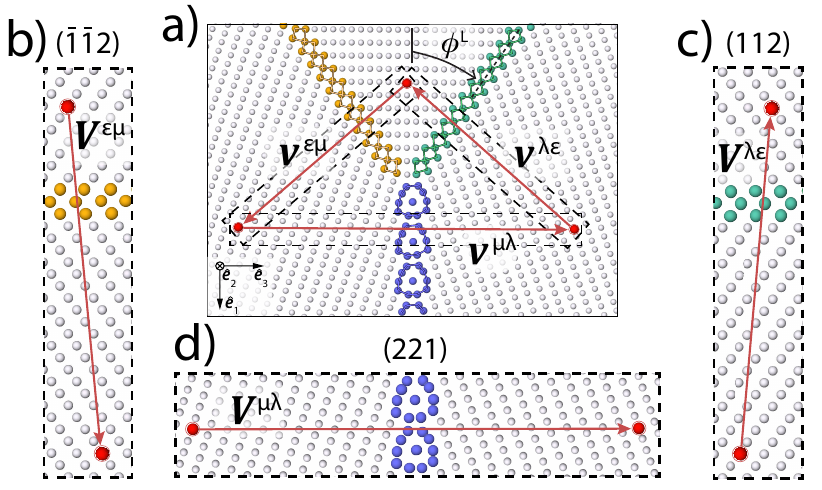}
\caption{Burgers vector calculation of a GB triple junction, consisting of intersecting $\Sigma 9 (221)[1\bar{1}0]$,  $\Sigma 3 (112)[1\bar{1}0]$ and  $\Sigma 3 (\bar{1}\bar{1}2)[1\bar{1}0]$ GBs. Panel a depicts the triple junction and the Burgers circuit. Panel b shows the mapping of the vector $\bm{v}^{\varepsilon\mu}$ to the reference system, which contains the $\Sigma 3 (\bar{1}\bar{1}2)[1\bar{1}0]$ GB only. Panel c shows the mapping of the vector $\bm{v}^{\lambda\varepsilon}$ to the reference system, which contains the $\Sigma 3 (112)[1\bar{1}0]$ GB only. Panel d shows the mapping of the vector $\bm{v}^{\mu\lambda}$ to the reference system, which contains the $\Sigma 9 (221)[1\bar{1}0]$ GB only.} 
\label{fig:triple_junction_example}
\end{figure}

To better understand how the twin is growing, we calculate the Burgers content of the triple junction as a function of position, an example calculation of which is shown in Fig. \ref{fig:triple_junction_example}. For a given TJ position we define a Burgers circuit in the current configuration, which consists of three crossing vectors: $\bm{v}^{\mu\lambda}$, $\bm{v}^{\lambda\varepsilon}$ and $\bm{v}^{\varepsilon\mu}$ and then map these crossing vectors into reference configurations. By applying Eq. \eqref{eq:tj-circuit} to the drawn Burgers circuit, the Burgers vector of the junction can be found. For ease of calculation the reference structures used in this treatment are all 0 K structures. This will introduce errors as $\bm{t}^{WS}$ is a function of temperature. However, because the structure seen at temperature belongs to the same GB phase as the 0 K reference structures, we expect the errors to be small.

\begin{figure}[ht!]
\centering
\includegraphics[scale=1]{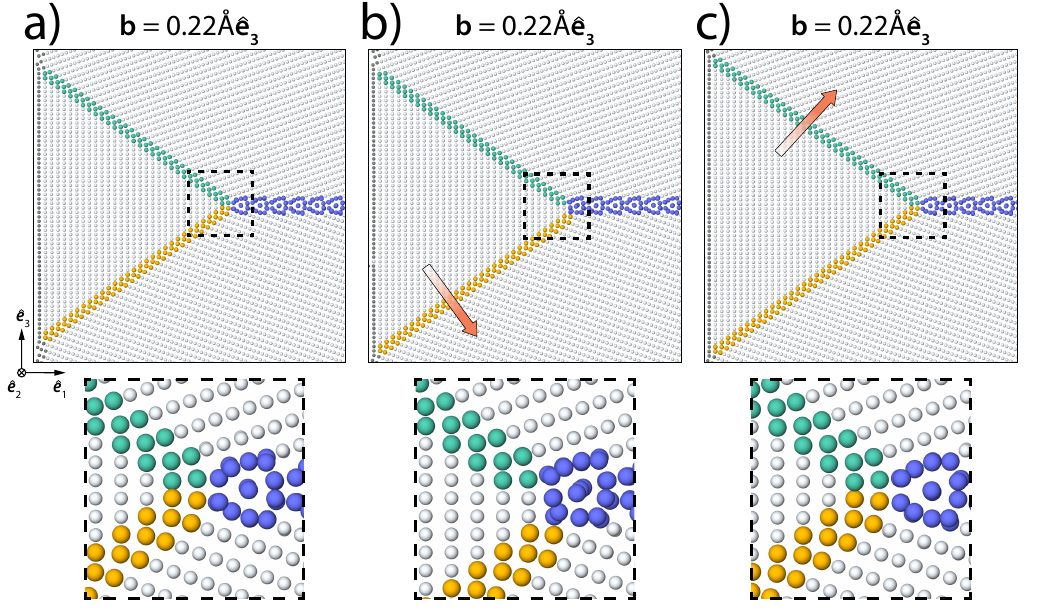}
\caption{ Snapshots of twin growth, with insets showing cyclic changes in the core structure of the TJ. The arrows in panels b and c indicate the order and direction of twin boundary motion. Panels a and c exhibit the same core structure, while panel b exhibits a different structure, due to nucleation and emission of a step. The TJ core transformation cycle is complete after the TJ emits two steps, one along each twin boundary.} 
\label{fig:changing_core_structure}
\end{figure}

As shown in supplemental video 1, the twin grows by discrete events, snapshots of which are given in Fig. \ref{fig:changing_core_structure}. The general mechanism of growth is that one twin boundary will move (Fig \ref{fig:changing_core_structure}b), growing the twin in the process. The movement of the twin boundary leads to a change in the core structure of the TJ. The other twin boundary will then move, causing a return to the original TJ core structure. Performing Burgers circuit analysis on these three configurations, we find that all three snapshots contain the same Burgers content: $\bm{b}^{TJ}=0.22\ \mathrm{\AA}\hat{\bm{e}}_3$. This means that for a given Burgers vector, multiple core structures are possible. The multiplicity of core structures can be understood within the context of the zero displacement incompletion condition \cite{PhysRevLett.119.246101,Thomas2019,SISANBAEV19923349}, which states that all GBs must meet at a single point to form a TJ. While at the atomic level the exact definition of the GB plane is, to a certain extent, arbitrary, it is still a useful framework for understanding TJ motion. If only one GB moves, as is the case in Fig. \ref{fig:changing_core_structure}b, we can think of the zero displacement incompletion condition as being violated. However, atomic rearrangement can occur at the TJ's core to accommodate the incompatibility between the positions of the three GBs leading to a new core structure. In other words, the incompatibility is accommodated by TJ core spreading. Such a picture suggests that we can use the displacement incompletion condition as a new metric for the core structure space of a TJ.

That the Burgers vector of the TJ does not change as the twin grows, suggests that twin growth is mediated by pure steps. For most growth events we were able to resolve the line defects mediating boundary motion. We find that twin boundary migration always begins at the TJ, not the free surface, and that for larger twin sizes all twin boundary motion events are mediated by pure steps, an example of which is given in Fig. \ref{fig:line-defect-examples}. As evidenced in supplemental video 1, we do not observe line defect motion along the $\Sigma 9$ GB directly contributing to twin growth in these simulations. We do not see large fluctuations in the step height of the $\Sigma 9$ GB that can be attributed to disconnections. We cannot state that the pure step is the only mode of twin growth, as we do not capture all growth events in detail due to our dump rate of 20 snapshots per ns. 
To confirm the absence of shear deformation during twin boundary migration, we have selected a region of marker atoms colored in gray, as shown in Fig. \ref{fig:line-defect-examples}, containing two 110 planes in the initial configuration prior to twin nucleation. In the case of shear-coupled motion \cite{CAHN20064953}, these marker planes would be expected to mirror across the twin boundary plane as the twin propagates. However, as shown in supplemental video 2 and Fig. \ref{fig:line-defect-examples}, after the twin boundary swept through a portion of the bicrystal, the marker planes remained aligned with their original positions, confirming the absence of shear-coupled motion during twin growth, confirming that glissile disconnections with finite Burgers vector are not involved in the twin growth process. 

\begin{figure}[h!]
\centering
\includegraphics[scale=0.8]{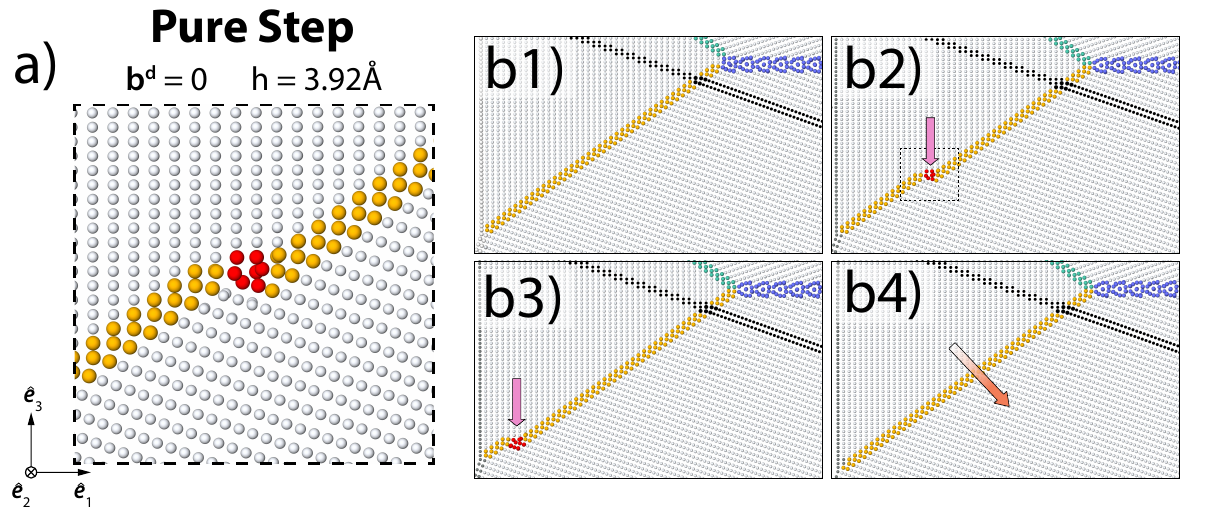}
\caption{Twin growth is mediated by the nucleation of pure steps from the GB TJ and their propagation to the free surface.  Panel a: a magnified view of the step's structure. Panel b: the propagation of the step from the TJ to the surface, resulting in twin boundary migration without shear. The dark gray atomic column marker are presented to show that no shear occurs after the passage of the steps during the entire twin growth process.} 
\label{fig:line-defect-examples}
\end{figure}  

\begin{table}
\caption{\label{tab:tj_221_112} The relevant parameters for calculating the Burgers vector associated with the triple junction illustrated in Fig.\ \ref{fig:triple_junction_example}. }
\begin{ruledtabular}
\begin{tabular}{lllllll}
GB & $\gamma\ \mathrm{(J/m^2)}$ & $t_1^{WS}$ ($\mathrm{\AA}$) & $t_2^{WS}$ ($\mathrm{\AA}$)  & $t_3^{WS}$ ($\mathrm{\AA}$) & $[n]$ \\
\hline
$\Sigma 9 (221)[1\bar{1}0]$ & 2.41 & 0.00 & 0.00 & -0.05 & $1/2$\\
$\Sigma 3 (112)[1\bar{1}0]$ & 0.58 & 0.00 & 0.00 & 0.10 &  0  
\end{tabular}
\end{ruledtabular}
\end{table}

Using Eq. \eqref{eq:b_tj}, we can calculate the smallest possible Burgers vector at the triple junction from the $\bm{t}^{WS}$ vectors of the constituent grain boundaries, given in TABLE \ref{tab:tj_221_112}, and the DSCL associated with the triple junction. Unlike in the previous example, the $\Sigma 9 (221) [1\bar{1}0]$ GB in question has a non-integer GB atomic fraction: $[n] = 1/2$. However, as shown in Ref. \cite{WINTER2025120968}, this change in excess number of atoms is directly incorporated into $\bm{t}^{WS}$, and as a such, Eq. \eqref{eq:b_tj}. The shear components of $\bm{t}^{WS}$ are zero for both boundaries. $t_3^{WS}$ of the twin boundary corresponds exactly to its excess volume $[V]_N=0.1$ $\mathrm{\AA}$. For the $\Sigma 9 (221) [1\bar{1}0]$ $t_3^{WS}=[V]_N+[n]d_{442}-d_{442}=-0.05$ $\mathrm{\AA}$ with $[V]$=0.211 $\mathrm{\AA}$ and $d_{442}=a/6=0.53$ $\mathrm{\AA}$ Due to the setup of this triple junction (see values from TABLE \ref{tab:tj_221_112})  Eq.\ \eqref{eq:b_tj} simplifies to 

\begin{subequations}\label{eq:tj_burgers_221_112}
\begin{align}
    b^{MDF}_1 &= -t_1^{WS,\mu\lambda},\\
    b^{MDF}_2 &= 0,\\
    b^{MDF}_3 &= -t_3^{WS,\mu\lambda} + 2t_3^{WS,\lambda\varepsilon} \cos \phi^L,
\end{align}
\end{subequations}


\noindent with $\bm{b}^{MDF} = 0.22 \ \mathrm{\AA} \hat{\bm{e}}_3$ for this system. 
Alternatively, the Burgers vector components can also be expressed in terms of the differences in the excess volumes $[V]_N$ and number of atoms $[n]$ of the constituent boundaries using the expressions for $t^{WS}$ given above. We note that the predicted Burgers vector is the same as that given by direct circuit mapping up to the significant figures displayed. This means that any disconnection emitted or absorbed by the TJ will increase its Burgers vector. By Frank's rule, this suggests that any disconnection emission event involving a finite Burgers vector will carry an energetic cost to it.

To illustrate this point, we can perform elasticity theory analysis to estimate the nucleation energy of a hypothetical glissile disconnection from the TJ. Importantly, this trichromatic pattern associated with the TJ forms a $\Sigma 9$ CSL. For this system the possible disconnections can be linear integer combinations of the DSCL primitive lattice vectors, which for the reference frame chosen in Fig. \ref{fig:triple_junction_example}, can be given as
\begin{subequations}
    \begin{align}
        \bm{a}^{DSC} &= \frac{a_0}{3\sqrt{2}}\hat{\bm{e}}_1+\frac{a_0}{6}\hat{\bm{e}}_3,\\
        \bm{b}^{DSC} &= \frac{a_0}{3\sqrt{2}}\hat{\bm{e}}_1-\frac{a_0}{6}\hat{\bm{e}}_3,\\
        \bm{c}^{DSC} &= \frac{a_0}{\sqrt{2}}\hat{\bm{e}}_2+\frac{a_0}{6}\hat{\bm{e}}_3.
    \end{align}
\end{subequations}

Ignoring image effects, we can estimate the elastic interaction per unit line length between an emitted glissile disconnection with Burgers vector, $\bm{b}^d$, and the TJ, with Burgers vector $\bm{b}^{TJ}-\bm{b}^d$, as

\begin{equation}\label{eq:dislocation-interactions}
    E^d/L= -\mu\left[\frac{(\bm{b}^d\cdot\bm{\xi})(\Delta\bm{b}^{TJ}\cdot\bm{\xi})}{2\pi}+\frac{(\bm{b}^d\times\bm{\xi})\cdot(\Delta\bm{b}^{TJ}\times\bm{\xi})}{2\pi(1-\nu)}\right]\ln\left(\frac{l}{\rho}\right)-\mu\frac{[(\bm{b}^d\times\bm{\xi})\cdot\bm{l}][(\Delta\bm{b}^{TJ}\times\bm{\xi})\cdot\bm{l}]}{2\pi(1-\nu)l^2},
\end{equation}
where $\Delta\bm{b}^{TJ}=\bm{b}^{TJ}-\bm{b}^d$ and $\rho$ is the core radius of the two dislocations. We approximate the shear modulus and Poisson ratio as $\mu=160$ GPa and $\nu=0.28$, the 0 K isotropic elastic constants for this potential and assume that $\rho=\Delta b^{TJ}$. To gain an order of magnitude estimate of the importance of elastic interactions on the energy barrier to twin boundary motion, and ignoring image effects in the process, we use Eq. \eqref{eq:dislocation-interactions} to consider the work done to move the disconnection 80 \AA, the approximate length of the twin boundary between the free surface and TJ shown in Fig. \ref{fig:changing_core_structure}a. We choose $\bm{b}^d=\bm{a}^{DSC}\approx 0.75\mathrm{\AA} \hat{\bm{e}}_1+0.53\mathrm{\AA}\hat{\bm{e}}_3$, because it gives the smallest possible Burgers vector for the triple junction which still allows for a disconnection to be emitted: $\Delta\bm{b}^{TJ}\approx 0.75\mathrm{\AA} \hat{\bm{e}}_1-0.31\mathrm{\AA}\hat{\bm{e}}_3$.

We find that for $l\approx9\ \mathrm{\AA}$, as is the case for this simulation cell, the change in elastic energy is 6 eV, which is much higher than  thermal fluctuations at 2500 K, $k_BT\approx0.2$ eV. As this analysis is meant to gain a qualitative understanding of the important energetics associated with twin growth in this simulation, we are neglecting any line force effects originating from the TJ. We do so as previous work has shown line forces to play an important role only when exceedingly small dislocation content is present at an interface \cite{PhysRevLett.90.246102,Winter2024,PhysRevMaterials.8.063606}. Such approximate analysis suggests that a large barrier to disconnection mediated twin growth exists, making it unlikely for disconnections, and shear coupled motion, to occur without some other driving force. In contrast, the energetic cost of the pure step in this analysis is zero. Of course, there is an energy barrier to the emission and propagation of the pure step, as evidenced by the discrete growth of the twin shown in this work. However, the qualitative elasticity analysis suggests that whatever barrier that exists is much lower than that associated with a disconnection with finite Burgers vector.

\section{Discussion}\label{sec:Discussion}

TJs are present in all polycrystalline microstructures and have long been recognized for their potential impact on diffusivity, impurity segregation, grain boundary migration, and microstructural stability. In these studies, TJs are typically treated as geometrical line defects characterized by excess properties localized to their core, rather than as dislocation-type defects with an associated long-range elastic field. Here, we derive the Burgers vector of a TJ and show that it can be decomposed into two contributions: a discrete DSCL component and an intrinsic component arising from excess properties of the adjoining grain boundaries captured by the $\bm{t}^{WS}$ vector. For regular GB disconnections on a flat GB there is no intrinsic contribution to the Burgers vector, because the $\bm{t}^{WS}$ vectors cancel each other: the circuit crosses identical GB structures resulting in a pure DSCL Burgers vector. For TJs, facet junctions and GB phase junctions these contributions from distinct interfacial structures will not cancel in general, leading to a non-zero non-DSCL component of the Burgers vector. As a result, TJs associated with a DSCL will possess an inherent dislocation character, with a finite Burgers vector that cannot be reduced to zero by absorbing regular GB disconnections.

This TJ property has important implications for microstructural evolution. Due to their intrinsic Burgers content TJs are expected to generate long-range elastic fields, and interact elastically with other defects, such as lattice dislocations, GB disconnections, other junctions and free surfaces influencing defect evolution during processes such as grain growth, plastic deformation, radiation damage and creep \cite{Shen1,Shen2,LI2025121364,Thomas2019,LIN201563}. Solute segregation to TJs should not be limited to excess solute at the junction core, but should also include the associated Cottrell atmosphere.

In addition to the intrinsic component, the Burgers vector of a TJ is defined up to a DSCL vector of the tricrystal that forms a CSL, implying that multiple configurations with different Burgers vectors and core structures may exist. Transitions between these configurations, mediated by the absorption or emission of point defects or interfacial defects, provide a mechanism for TJ activity during microstructural evolution. In this context, TJs may act as preferred sites for defect nucleation, with their behavior governed not only by the lowest-energy configuration but also by a spectrum of metastable states. Therefore, the energetics associated with these states are expected to play an important role in processes involving defect reactions during interface motion and deformation.

Using molecular dynamics simulations, we examine the nucleation and growth of a new grain at a free surface terminating a high-angle, high-energy grain boundary. The nucleation process is driven by a reduction in the total interfacial energy, as the emerging grain is bounded by a closest packed, low-energy surface and two low-energy coherent twin boundaries. The growth of the newly formed grain reveals two interesting aspects. First, growth proceeds entirely through the nucleation of pure steps at a single grain boundary TJ. The new grain is bounded by three distinct TJs: one grain boundary TJ and two symmetry-equivalent surface TJs. Although surface TJs can serve as nucleation sites due to the driving force associated with the formation of the low-energy $\langle110\rangle$ surface, remarkably, no nucleation activity is observed at the surface TJs. This observation indicates that not all TJs are equivalent in their ability to nucleate interfacial defects which may depend on their local structure, energetics and the driving force. Some TJs may exhibit significantly lower disconnection activation barriers and may play a more dominant role in microstructural evolution. Future work should explore the nucleation properties of different grain boundary TJs and its relation to crystallography and driving force, especially in light of recent studies by Barnett at al demonstrating that solute segregation can strongly modify this TJ behavior\cite{Barnett2024}.

Second, the twinned grain is found to grow without producing shear, in contrast to deformation twinning mechanisms. During growth the coherent twin boundaries do not move by coupled motion that involves relative grain translations parallel to the boundary plane \cite{CAHN20064953}, and advance instead by passing pure steps. The fact that the twinned grain grows from the TJ bears some semblance to the TJ-mediated mechanism of annealing twin nucleation \cite{LIN201563}, with the important difference that in our case the TJ forms during the nucleation of the twin from the free surface, while in the case of Ref. \cite{LIN201563} the twin nucleates from a preexisting TJ.

We observe no non-zero Burgers vector disconnection nucleation events that could generate shear. This observation can be understood from Frank's rule; the elastic energy of a dislocation scales with $b^2$. As such, a TJ that does not have the smallest possible Burgers vector will result in a higher energy within the system.  As shown above, such an emission event experiences its own energetic barrier due to the elastic interaction between the emitted disconnection and residual dislocation content at the TJ.  This barrier is especially important in the system studied where there is no externally applied stress present to overcome the barrier. These observations suggest that step-mediated growth provides a lower-energy pathway under the present conditions, further underscoring the central role of triple junctions in governing interfacial kinetics.


In this work, we have developed a method to quantify the dislocation content existing at grain boundary junctions using a Burgers circuit mapping technique. We have shown that for a general GB junction that is associated with a DSCL, the possible dislocation content at the junction can be calculated in terms of the individual excess properties of the grain boundaries that make up the junction. We have used this approach to quantify dislocation content in a facet junction and triple junction in atomistic simulations of tungsten. We have further used this methodology to identify the mechanisms by which a twin can nucleate from the intersection of a free surface and a high angle grain boundary.

\section{Acknowledgments}

This work was performed under the auspices of the U.S. Department of Energy (DOE) by Lawrence Livermore National Laboratory under contract DE-AC52-07NA27344. TF was supported by the U.S. DOE, Office of Science under an Office of Fusion Energy Sciences Early Career Award. RER was supported by the U.S. DOE, Office of Science, Office of Fusion Energy Sciences. Computing support for this work came from the Lawrence Livermore National Laboratory Institutional Computing Grand Challenge program. An award of computer time was also provided by the INCITE program. This research used resources of both the Argonne and Oak Ridge Leadership Computing Facilities, which are DOE Office of Science User Facilities supported under contracts DE-AC02-06CH11357 and DE-AC05-00OR22725, respectively. ISW and RDM received support from the LDRD program at Sandia National Laboratories. Sandia National Laboratories is a multi-mission laboratory managed and operated by National Technology \& Engineering Solutions of Sandia, LLC (NTESS), a wholly owned subsidiary of Honeywell International Inc., for the U.S. Department of Energy’s National Nuclear Security Administration (DOE/NNSA) under contract DE-NA0003525.


\bibliography{aapmsamp}

@PREAMBLE{
 "\providecommand{\noopsort}[1]{}" 
 # "\providecommand{\singleletter}[1]{#1}%" 
}

@article{stukowski2010visualization,
  title={Visualization and analysis of atomistic simulation data with OVITO--the Open Visualization Tool},
  author={Stukowski, Alexander},
  journal={Modelling and simulation in materials science and engineering},
  volume={18},
  number={1},
  pages={015012},
  year={2010}
}

@article{ADMAL2022118340,
title = {Interface dislocations and grain boundary disconnections using Smith normal bicrystallography},
journal = {Acta Materialia},
volume = {240},
pages = {118340},
year = {2022},
issn = {1359-6454},
doi = {https://doi.org/10.1016/j.actamat.2022.118340},
url = {https://www.sciencedirect.com/science/article/pii/S1359645422007194},
author = {Nikhil Chandra Admal and Tusher Ahmed and Enrique Martinez and Giacomo Po}
}

@article{Plimpton1995Fast,
title = {Fast Parallel Algorithms for Short-Range Molecular Dynamics},
journal = {Journal of Computational Physics},
volume = {117},
number = {1},
pages = {1-19},
year = {1995},
issn = {0021-9991},
doi = {https://doi.org/10.1006/jcph.1995.1039},
url = {https://www.sciencedirect.com/science/article/pii/S002199918571039X},
author = {Steve Plimpton}
}

@article{brunger1984stochastic,
  title={Stochastic boundary conditions for molecular dynamics simulations of ST2 water},
  author={Br{\"u}nger, Axel and Brooks III, Charles L and Karplus, Martin},
  journal={Chemical physics letters},
  volume={105},
  number={5},
  pages={495--500},
  year={1984},
  publisher={Elsevier}
}

@BOOK{HirthLothe,
   author       = "J. P. Hirth and J. Lothe", 
   title        = "Theory of Dislocations", 
   publisher    = "Krieger", 
   year         = "1982", 
   edition      = "2", 
   address      = "Malabar, Florida", 
}

@incollection{POND1994287,
title = {Defects at Surfaces and Interfaces},
editor = {HENRY EHRENREICH and DAVID TURNBULL},
series = {Solid State Physics},
publisher = {Academic Press},
volume = {47},
pages = {287-365},
year = {1994},
issn = {0081-1947},
doi = {https://doi.org/10.1016/S0081-1947(08)60641-4},
url = {https://www.sciencedirect.com/science/article/pii/S0081194708606414},
author = {R.C. Pond and J.P. Hirth}
}

@article {Thomas2019,
	author = {Thomas, Spencer L. and Wei, Chaozhen and Han, Jian and Xiang, Yang and Srolovitz, David J.},
	title = {Disconnection description of triple-junction motion},
	volume = {116},
	number = {18},
	pages = {8756--8765},
	year = {2019},
	doi = {10.1073/pnas.1820789116},
	publisher = {National Academy of Sciences},
	issn = {0027-8424},
	URL = {https://www.pnas.org/content/116/18/8756},
	journal = {Proceedings of the National Academy of Sciences}
}

@article{PhysRevLett.119.246101,
  title = {Equation of Motion for a Grain Boundary},
  author = {Zhang, Luchan and Han, Jian and Xiang, Yang and Srolovitz, David J.},
  journal = {Phys. Rev. Lett.},
  volume = {119},
  issue = {24},
  pages = {246101},
  numpages = {5},
  year = {2017},
  month = {Dec},
  publisher = {American Physical Society},
  doi = {10.1103/PhysRevLett.119.246101},
  url = {https://link.aps.org/doi/10.1103/PhysRevLett.119.246101}
}

@article{HAN2022117178,
title = {Disconnection-mediated migration of interfaces in microstructures: I. continuum model},
journal = {Acta Materialia},
volume = {227},
pages = {117178},
year = {2022},
issn = {1359-6454},
doi = {https://doi.org/10.1016/j.actamat.2021.117178},
url = {https://www.sciencedirect.com/science/article/pii/S1359645421005589},
author = {Jian Han and David J. Srolovitz and Marco Salvalaglio}
}

@article{SRINIVASAN19992821,
title = {Excess energy of grain-boundary trijunctions: an atomistic simulation study},
journal = {Acta Materialia},
volume = {47},
number = {9},
pages = {2821-2829},
year = {1999},
issn = {1359-6454},
doi = {https://doi.org/10.1016/S1359-6454(99)00120-2},
url = {https://www.sciencedirect.com/science/article/pii/S1359645499001202},
author = {S. G. Srinivasan and J. W. Cahn and H. Jónsson and G. Kalonji}
}

@article{seki2023incommensurate,
  title={Incommensurate grain-boundary atomic structure},
  author={Seki, Takehito and Futazuka, Toshihiro and Morishige, Nobusato and Matsubara, Ryo and Ikuhara, Yuichi and Shibata, Naoya},
  journal={Nature Communications},
  volume={14},
  number={1},
  pages={7806},
  year={2023},
  publisher={Nature Publishing Group UK London}
}

@article{CAHN20064953,
title = {Coupling grain boundary motion to shear deformation},
journal = {Acta Materialia},
volume = {54},
number = {19},
pages = {4953-4975},
year = {2006},
issn = {1359-6454},
doi = {https://doi.org/10.1016/j.actamat.2006.08.004},
url = {https://www.sciencedirect.com/science/article/pii/S1359645406005313},
author = {John W. Cahn and Yuri Mishin and Akira Suzuki}
}

@article{LIN201563,
title = {Observation of annealing twin nucleation at triple lines in nickel during grain growth},
journal = {Acta Materialia},
volume = {99},
pages = {63-68},
year = {2015},
issn = {1359-6454},
doi = {https://doi.org/10.1016/j.actamat.2015.07.041},
url = {https://www.sciencedirect.com/science/article/pii/S1359645415005121},
author = {B. Lin and Y. Jin and C.M. Hefferan and S.F. Li and J. Lind and R.M. Suter and M. Bernacki and N. Bozzolo and A.D. Rollett and G.S. Rohrer}
}

@article{LI2025121364,
title = {Line and planar defects with zero formation free energy: Applications of the phase rule towards ripening-immune microstructures},
journal = {Acta Materialia},
volume = {298},
pages = {121364},
year = {2025},
issn = {1359-6454},
doi = {https://doi.org/10.1016/j.actamat.2025.121364},
url = {https://www.sciencedirect.com/science/article/pii/S1359645425006500},
author = {Ju Li and Yuri Mishin}
}

@article{King01052007,
author = {A. H. King},
title = {Triple junction energy and prospects for measuring it},
journal = {Materials Science and Technology},
volume = {23},
number = {5},
pages = {505--508},
year = {2007},
publisher = {Taylor \& Francis},
doi = {10.1179/174328407X176910},
}

@article{Yin2003,
    author = {Yin, K-M and King, A H and Hsieh, TE and Chen, F-R and Kai, J J and Chang, L},
    title = {Segregation of Bismuth to Triple Junctions in Copper},
    journal = {Microscopy and Microanalysis},
    volume = {3},
    number = {5},
    pages = {417-422},
    year = {2003},
    month = {01},
    issn = {1431-9276},
    doi = {10.1017/S1431927697970318},
}

@Article{Upmanyu1999,
author="Upmanyu, Moneesh
and Srolovitz, D. J.
and Shvindlerman, L. S.
and Gottstein, G.",
title="Triple Junction Mobility: A Molecular Dynamics Study",
journal="Interface Science",
year="1999",
month="Nov",
day="01",
volume="7",
number="3",
pages="307--319",
issn="1573-2746",
doi="10.1023/A:1008781611991",
url="https://doi.org/10.1023/A:1008781611991"
}

@Article{Gottstein1999,
author="Gottstein, G{\"u}nter
and Sursaeva, Vera
and Shvindlerman, Lasar S.",
title="The Effect of Triple Junctions on Grain Boundary Motion and Grain Microstructure Evolution",
journal="Interface Science",
year="1999",
month="Nov",
day="01",
volume="7",
number="3",
pages="273--283",
issn="1573-2746",
doi="10.1023/A:1008721426104",
url="https://doi.org/10.1023/A:1008721426104"
}

@article{GOTTSTEIN2000397,
title = "The effect of triple-junction drag on grain growth",
journal = "Acta Materialia",
volume = "48",
number = "2",
pages = "397 - 403",
year = "2000",
issn = "1359-6454",
doi = "https://doi.org/10.1016/S1359-6454(99)00373-0",
url = "http://www.sciencedirect.com/science/article/pii/S1359645499003730",
author = "G Gottstein and A. H King and L. S Shvindlerman"
}

@article{GOTTSTEIN2002703,
title = "Triple junction drag and grain growth in 2D polycrystals",
journal = "Acta Materialia",
volume = "50",
number = "4",
pages = "703 - 713",
year = "2002",
issn = "1359-6454",
doi = "https://doi.org/10.1016/S1359-6454(01)00391-3",
url = "http://www.sciencedirect.com/science/article/pii/S1359645401003913",
author = "G. Gottstein and L. S. Shvindlerman"
}

@article{CZUBAYKO19985863,
title = "Influence of triple junctions on grain boundary motion",
journal = "Acta Materialia",
volume = "46",
number = "16",
pages = "5863 - 5871",
year = "1998",
issn = "1359-6454",
doi = "https://doi.org/10.1016/S1359-6454(98)00241-9",
url = "http://www.sciencedirect.com/science/article/pii/S1359645498002419",
author = "U. Czubayko and V. G. Sursaeva and G. Gottstein and L. S. Shvindlerman"
}

@article{PhysRevLett.90.246102,
  title = {Why Do Grain Boundaries Exhibit Finite Facet Lengths?},
  author = {Hamilton, J. C. and Siegel, Donald J. and Daruka, Istvan and L\'eonard, Fran\ifmmode \mbox{\c{c}}\else \c{c}\fi{}ois},
  journal = {Phys. Rev. Lett.},
  volume = {90},
  issue = {24},
  pages = {246102},
  numpages = {4},
  year = {2003},
  month = {Jun},
  publisher = {American Physical Society},
  doi = {10.1103/PhysRevLett.90.246102},
  url = {https://link.aps.org/doi/10.1103/PhysRevLett.90.246102}
}

@article{Barnett2024,
author = {Barnett, Annie K. and Hussein, Omar and Alghalayini, Maher and Hinojos, Alejandro and Nathaniel, James E. II and Medlin, Douglas L. and Hattar, Khalid and Boyce, Brad L. and Abdeljawad, Fadi},
title = {Triple Junction Segregation Dominates the Stability of Nanocrystalline Alloys},
journal = {Nano Letters},
volume = {24},
number = {31},
pages = {9627-9634},
year = {2024},
doi = {10.1021/acs.nanolett.4c02395},
}

@article{PENG2022117522,
title = {Quantitative analysis of grain boundary diffusion, segregation and precipitation at a sub-nanometer scale},
journal = {Acta Materialia},
volume = {225},
pages = {117522},
year = {2022},
issn = {1359-6454},
doi = {https://doi.org/10.1016/j.actamat.2021.117522},
url = {https://www.sciencedirect.com/science/article/pii/S1359645421009009},
author = {Zirong Peng and Thorsten Meiners and Yifeng Lu and Christian H. Liebscher and Aleksander Kostka and Dierk Raabe and Baptiste Gault}
}

@article{dimitrakopulos1997defect,
  title={The defect character of interface junction lines},
  author={Dimitrakopulos, G P and Karakostas, T H and Pond, R C},
  journal={Interface Science},
  volume={4},
  number={1},
  pages={129--138},
  year={1997},
  publisher={Springer},
  doi={10.1007/BF00200843}
}

@article{REDACHELLALI2013164,
title = {Nano-analysis of grain boundary and triple junction transport in nanocrystalline Ni/Cu},
journal = {Ultramicroscopy},
volume = {132},
pages = {164-170},
year = {2013},
note = {IFES 2012},
issn = {0304-3991},
doi = {https://doi.org/10.1016/j.ultramic.2012.12.002},
url = {https://www.sciencedirect.com/science/article/pii/S0304399112002896},
author = {Mohammed {Reda Chellali} and Zoltan Balogh and Guido Schmitz}
}

@article{ZHAO20105646,
title = {Measurement of grain boundary triple line energy in copper},
journal = {Acta Materialia},
volume = {58},
number = {17},
pages = {5646-5653},
year = {2010},
issn = {1359-6454},
doi = {https://doi.org/10.1016/j.actamat.2010.06.039},
author = {B. Zhao and J. Ch. Verhasselt and L. S. Shvindlerman and G. Gottstein}
}

@book{bollmann2012crystal,
  title={{C}rystal {D}efects and {C}rystalline {I}nterfaces},
  author={Bollmann, Walter},
  year={2012},
  publisher={Springer Science \& Business Media, Berlin}
}

@incollection{Herring,
    booktitle={The Physics of Powder Metallurgy},
    title={},
    author={Herring, Conyers},
    publisher = {NY.: McGraw-Hill},
    pages={143},
    year = {1951}
}

@article{KIM20093662,
title = {Anomalous triple junction surface pits in nanocrystalline zirconia thin films and their relationship to triple junction energy},
journal = {Acta Materialia},
volume = {57},
number = {12},
pages = {3662-3670},
year = {2009},
issn = {1359-6454},
doi = {https://doi.org/10.1016/j.actamat.2009.04.032},
url = {https://www.sciencedirect.com/science/article/pii/S1359645409002560},
author = {Hakkwan Kim and Yi Xuan and Peide D. Ye and Raghavan Narayanan and Alexander H. King}
}

@article{FORTIER1991177,
title = {Triple line energy determination by scanning tunneling microscopy},
journal = {Scripta Metallurgica et Materialia},
volume = {25},
number = {1},
pages = {177-182},
year = {1991},
issn = {0956-716X},
doi = {https://doi.org/10.1016/0956-716X(91)90376-C},
url = {https://www.sciencedirect.com/science/article/pii/0956716X9190376C},
author = {P Fortier and G Palumbo and G. D Bruce and W. A Miller and K. T Aust}
}

@article{UPMANYU20021405,
title = {Molecular dynamics simulation of triple junction migration},
journal = {Acta Materialia},
volume = {50},
number = {6},
pages = {1405-1420},
year = {2002},
issn = {1359-6454},
doi = {https://doi.org/10.1016/S1359-6454(01)00446-3},
url = {https://www.sciencedirect.com/science/article/pii/S1359645401004463},
author = {M Upmanyu and D. J Srolovitz and L. S Shvindlerman and G Gottstein}
}

@article{Winter2024,
  title = {Phase Pattern Formation in Grain Boundaries},
  author = {Winter, I. S. and Frolov, T.},
  journal = {Phys. Rev. Lett.},
  volume = {132},
  issue = {18},
  pages = {186204},
  numpages = {6},
  year = {2024},
  month = {May},
  publisher = {American Physical Society},
  doi = {10.1103/PhysRevLett.132.186204},
  url = {https://link.aps.org/doi/10.1103/PhysRevLett.132.186204}
}

@article{WINTER2025120968,
title = {Quantifying and visualizing the microscopic degrees of freedom of grain boundaries in the Wigner–Seitz cell of the displacement-shift-complete lattice},
journal = {Acta Materialia},
volume = {291},
pages = {120968},
year = {2025},
issn = {1359-6454},
doi = {https://doi.org/10.1016/j.actamat.2025.120968},
url = {https://www.sciencedirect.com/science/article/pii/S1359645425002599},
author = {I. S. Winter and T. Frolov}
}

@article{upmanyu1999triple,
  title={Triple junction mobility: A molecular dynamics study},
  author={Upmanyu, Moneesh and Srolovitz, David J and Shvindlerman, L S and Gottstein, G},
  journal={Interface Science},
  volume={7},
  number={3},
  pages={307--319},
  year={1999},
  publisher={Springer}
}

@article{PhysRevMaterials.8.063606,
  title = {Stable nanofacets in [111] tilt grain boundaries of face-centered cubic metals},
  author = {Brink, Tobias and Langenohl, Lena and Pemma, Swetha and Liebscher, Christian H. and Dehm, Gerhard},
  journal = {Phys. Rev. Mater.},
  volume = {8},
  issue = {6},
  pages = {063606},
  numpages = {10},
  year = {2024},
  month = {Jun},
  publisher = {American Physical Society},
  doi = {10.1103/PhysRevMaterials.8.063606},
  url = {https://link.aps.org/doi/10.1103/PhysRevMaterials.8.063606}
}

@article{MEDLIN2017383,
title = "Defect character at grain boundary facet junctions: Analysis of an asymmetric Σ = 5 grain boundary in Fe",
journal = "Acta Materialia",
volume = "124",
pages = "383 - 396",
year = "2017",
issn = "1359-6454",
doi = "https://doi.org/10.1016/j.actamat.2016.11.017",
url = "http://www.sciencedirect.com/science/article/pii/S1359645416308758",
author = "D. L. Medlin and K. Hattar and J. A. Zimmerman and F. Abdeljawad and S. M. Foiles"
}

@article{Frolov2013,
author = {Frolov, Timofey and Olmsted, David L and Asta, Mark and Mishin, Yuri},
doi = {10.1038/ncomms2919},
issn = {2041-1723},
journal = {Nature Communications},
number = {1},
pages = {1899},
title = {{Structural phase transformations in metallic grain boundaries}},
url = {https://doi.org/10.1038/ncomms2919},
volume = {4},
year = {2013}
}

@article{EICH2016364,
title = {Embedded-atom study of low-energy equilibrium triple junction structures and energies},
journal = {Acta Materialia},
volume = {109},
pages = {364-374},
year = {2016},
issn = {1359-6454},
doi = {https://doi.org/10.1016/j.actamat.2016.02.058},
url = {https://www.sciencedirect.com/science/article/pii/S135964541630129X},
author = {S.M. Eich and G. Schmitz}
}

@article{SISANBAEV19923349,
title = {The effect of triple junction type on grain-boundary sliding and accomodation in aluminium tricrystals},
journal = {Acta Metallurgica et Materialia},
volume = {40},
number = {12},
pages = {3349-3356},
year = {1992},
issn = {0956-7151},
doi = {https://doi.org/10.1016/0956-7151(92)90048-J},
url = {https://www.sciencedirect.com/science/article/pii/095671519290048J},
author = {A.V. Sisanbaev and R.Z. Valiev}
}

@article{PhysRevB.79.174110,
  title = {Molecular dynamics modeling of self-diffusion along a triple junction},
  author = {Frolov, T. and Mishin, Y.},
  journal = {Phys. Rev. B},
  volume = {79},
  issue = {17},
  pages = {174110},
  numpages = {5},
  year = {2009},
  month = {May},
  publisher = {American Physical Society},
  doi = {10.1103/PhysRevB.79.174110},
  url = {https://link.aps.org/doi/10.1103/PhysRevB.79.174110}
}

@article{FEDOROV200251,
title = {Triple junction diffusion and plastic flow in fine-grained materials},
journal = {Scripta Materialia},
volume = {47},
number = {1},
pages = {51-55},
year = {2002},
issn = {1359-6462},
doi = {https://doi.org/10.1016/S1359-6462(02)00096-9},
url = {https://www.sciencedirect.com/science/article/pii/S1359646202000969},
author = {A.A. Fedorov and M.Yu. Gutkin and I.A. Ovid'ko}
}

@article{GOTTSTEIN20051535,
title = {Triple junction motion and grain microstructure evolution},
journal = {Acta Materialia},
volume = {53},
number = {5},
pages = {1535-1544},
year = {2005},
issn = {1359-6454},
doi = {https://doi.org/10.1016/j.actamat.2004.12.006},
url = {https://www.sciencedirect.com/science/article/pii/S1359645404007426},
author = {G. Gottstein and Y. Ma and L.S. Shvindlerman}
}

@article{JOHNSON2014134,
title = {A phase-field model for grain growth with trijunction drag},
journal = {Acta Materialia},
volume = {67},
pages = {134-144},
year = {2014},
issn = {1359-6454},
doi = {https://doi.org/10.1016/j.actamat.2013.12.012},
url = {https://www.sciencedirect.com/science/article/pii/S1359645413009476},
author = {A.E. Johnson and P.W. Voorhees}
}

@article{CHEN2007253,
title = {Contribution of triple junctions to the diffusion anomaly in nanocrystalline materials},
journal = {Scripta Materialia},
volume = {57},
number = {3},
pages = {253-256},
year = {2007},
issn = {1359-6462},
doi = {https://doi.org/10.1016/j.scriptamat.2007.03.057},
url = {https://www.sciencedirect.com/science/article/pii/S1359646207002618},
author = {Ying Chen and Christopher A. Schuh}
}

@article{TUCHINDA2024120274,
title = {Triple junction excess energy in polycrystalline metals},
journal = {Acta Materialia},
volume = {279},
pages = {120274},
year = {2024},
issn = {1359-6454},
doi = {https://doi.org/10.1016/j.actamat.2024.120274},
url = {https://www.sciencedirect.com/science/article/pii/S1359645424006244},
author = {Nutth Tuchinda and Christopher A. Schuh}
}

@article{TUCHINDA2025121429,
title = {Triple junction solute segregation and the stability of nanocrystalline alloys},
journal = {Acta Materialia},
volume = {299},
pages = {121429},
year = {2025},
issn = {1359-6454},
doi = {https://doi.org/10.1016/j.actamat.2025.121429},
url = {https://www.sciencedirect.com/science/article/pii/S1359645425007153},
author = {Nutth Tuchinda and Yu-ning Chiu and Christopher A. Schuh}
}

@article{chen2024grand,
  title={Grand canonically optimized grain boundary phases in hexagonal close-packed titanium},
  author={Chen, Enze and Heo, Tae Wook and Wood, Brandon C and Asta, Mark and Frolov, Timofey},
  journal={Nature Communications},
  volume={15},
  number={1},
  pages={7049},
  year={2024},
  publisher={Nature Publishing Group UK London},
  doi={10.1038/s41467-024-51330-9}
}

@article{GRIMMER19741221,
title = "A reciprocity relation between the coincidence site lattice and the DSC lattice",
journal = "Scripta Metallurgica",
volume = "8",
number = "11",
pages = "1221 - 1223",
year = "1974",
issn = "0036-9748",
doi = "https://doi.org/10.1016/0036-9748(74)90334-2",
url = "http://www.sciencedirect.com/science/article/pii/0036974874903342",
author = "Hans Grimmer"
}

@article{ZHOU20014005,
title = {Atomic scale structure of sputtered metal multilayers},
journal = {Acta Materialia},
volume = {49},
number = {19},
pages = {4005-4015},
year = {2001},
issn = {1359-6454},
doi = {https://doi.org/10.1016/S1359-6454(01)00287-7},
url = {https://www.sciencedirect.com/science/article/pii/S1359645401002877},
author = {X. W. Zhou and H. N. G. Wadley and R. A. Johnson and D. J. Larson and N. Tabat and A. Cerezo and A. K. Petford-Long and G. D. W. Smith and P. H. Clifton and R. L. Martens and T. F. Kelly}
}

@article{Shen1,
title = {A nucleation rate limited model for grain boundary creep},
journal = {Acta Materialia},
volume = {246},
pages = {118718},
year = {2023},
issn = {1359-6454},
doi = {https://doi.org/10.1016/j.actamat.2023.118718},
url = {https://www.sciencedirect.com/science/article/pii/S1359645423000502},
author = {Shen J. Dillon and Eric Lang and Sarah C. Finkeldei and Jia-hu Ouyang and Khalid Hattar}
}

@article{Shen2,
title = {An Arrhenius mechanics model for polycrystalline plasticity and creep},
journal = {Acta Materialia},
volume = {305},
pages = {121831},
year = {2026},
issn = {1359-6454},
doi = {https://doi.org/10.1016/j.actamat.2025.121831},
url = {https://www.sciencedirect.com/science/article/pii/S1359645425011188},
author = {Ahmad Mirzaei and Shen J. Dillon}
}

@article{frolov_2018,
  title = {Grain boundary phases in bcc metals},
  author = {Frolov, Timofey and Setyawan, Wahyu and Kurtz, Richard J. and Marian, Jaime and Oganov, Artem E. and Rudd, Robert E. and Zhu, Qiang},
  year = {2018},
  journal = {Nanoscale},
  volume = {10},
  number = {17},
  pages = {8253--8268},
  doi = {10.1039/C8NR00271A}
}
\end{document}